\documentclass[doublecol]{epl2}
\usepackage{color}
\usepackage{graphicx}
\usepackage{dcolumn}
\usepackage{bm}
\usepackage{array}
\usepackage{amsmath}
\newcommand{\PreserveBackslash}[1]{\let\temp=\\#1\let\\=\temp}
\newcolumntype{C}[1]{>{\PreserveBackslash\centering}p{#1}}
\newcolumntype{R}[1]{>{\PreserveBackslash\raggedleft}p{#1}}
\newcolumntype{L}[1]{>{\PreserveBackslash\raggedright}p{#1}}
\begin{document}



\title{Predicting missing links and their weights via reliable-route-based method }

\author{Jing Zhao\footnote{Corresponding author: \email{zhaojanne@gmail.com}}\inst{1} \and Lili Miao\inst{2} \and Haiyang Fang\inst{1} \and Qian-Ming Zhang\inst{2,3} \and Min Nie\inst{2} \and Tao Zhou\footnote{Corresponding author: \email{zhutou@ustc.edu}}\inst{2,3}}
\shortauthor{JING ZHAO \etal}
\institute{
\inst{1}Department of Mathematics, Logistical Engineering University,
Chongqing 404100, People's Republic of China\\
\inst{2}
Web Sciences Center, School of Computer Science and Engineering, University of Electronic Science and Technology of China,
Chengdu 610054, People's Republic of China\\
\inst{3}
Big Data Research Center, University of Electronic Science and Technology of China,
Chengdu 611731, People's Republic of China}


\abstract
{Link prediction aims to uncover missing links or predict the emergence of future relationships according to the current networks structure. Plenty of algorithms have been developed for link prediction in unweighted networks, with only a very few of them having been extended to weighted networks. Thus far, how to predict weights of links is important but rarely studied. In this Letter, we present a reliable-route-based method to extend unweighted local similarity indices to weighted indices and propose a method to predict both the link existence and link weights accordingly. Experiments on different real networks suggest that the weighted resource allocation index has the best performance to predict the existence of links, while the reliable-route-based weighted resource allocation index performs noticeably better on weight prediction. Further analysis shows a strong correlation for both link prediction and weight prediction: the larger the clustering coefficient, the higher the prediction accuracy.}

\maketitle

\section{Introduction}
Link prediction aims to predict whether there exists a missing link between a pair of non-connected nodes or whether a link may appear in the future \cite{1}. Link prediction can not only help to find missing data in real networks, but also complement our understanding on the evolution processes of networks \cite{2,3}. Thus it has attracted increasing attentions in recent years. Ideally, topological features of the network and profiles of the considered nodes are combined to assess the possibility of the existence of links between nodes. For example, when predicting literature citations based on the citation network, Popescul and Ungar considered not only topological characteristics of the network, but also the node attributes, such as authors, journal names and contents of the papers\cite{4}. However, in most cases, it is very difficult to get attribute information of nodes or the information is not reliable. For example, in online social networks, the information of many users is confidential or false. Therefore, many algorithms only make use of topological information. A few variant problems make use of functional outputs \cite{Barzel2013} or predict link direction instead of link existence \cite{Guo2013}. Up to now, the mainstream of topological based methods is the similarity-based algorithm, including algorithms that use local, global or quasi-local information respectively to compute the similarity indices \cite{1,14}.

Most previous studies in link prediction focus on unweighted networks. Recently, a few works tried to extend similarity indices from unweighted networks to weighted networks \cite{5,6,7,8,9,Wind2012}. Some strategies have been proposed to enhance the precision of weighted indices, for example, to emphasize the contributions of weak links \cite{5}, or to integrate multiple indices, including local and global ones \cite{6}. However, these studies did not consider the prediction of weights, which is also very significant in addition to the prediction of the existence of links.

In this Letter, we try to predict missing links and their weights using local similarity measures. Inspired by the solution of the \emph{most reliable route problem} in communication networks \cite{10}, we propose a reliable-route-based method to generalize unweighted similarity indices to weighted ones. Considering that the similarity index between two unconnected nodes essentially reflects interaction strength between the two nodes, we set weights of missing links proportional to the similarity indices. We apply an optimization algorithm to identify the optimum weight prediction function and measure the accuracy of weight prediction by the Pearson correlation coefficient. Six real networks are used in experiments to evaluate our methods, indicating that the weighted resource allocation index and the reliable-route-based weighted resource allocation index outperform other indices in predicting the link existence and link weights, respectively.

\begin{table*}[!ht]
\caption{Basic topological features of the networks under study. $|V|$ and $|E|$ are the number of nodes and links, $\langle k\rangle$ is the average degree, and $C$ and $C_{w}$ are the unweighted and weighted versions (see Eq. (13) and Eq. (14) for definitions) of clustering coefficient, respectively.
\label{topology}}
\centering
{\begin{tabular}{cccccccc}
\hline
Networks &hsaPPi &Cel &Geom &String &Corum & String\_Corum\\
\hline
$|V|$    &2821 &281 &7343 &16886 &2314 &2270\\
$|E|$    &13880 &2402 &11898 &1520927 &34148 &153788\\
$\langle k\rangle$    &9.84 &17.1 &3.24 &180.14 &29.51 &135.49\\
$C$      &0.169 &0.346 &0.486 &0.301 &0.747 &0.371\\
$C_{w}$     &0.167 &0.291 &0.535 &0.231 &0.795 &0.278\\
\hline
\end{tabular}}
\end{table*}

\section{Problem and Metrics}
Given a weighted undirected network $G(V,E,W)$, where $V$, $E$ and $W$ are sets of nodes, links and link weights, respectively, we want to find out its missing links (or links that may appear in the future) and predict their weights as well. To do this, for each pair of unconnected nodes $x, y\in V$, we assign a likelihood score $S_{xy}$ to quantify the probability that ($x$, $y$) is a missing link. Then all unconnected pairs are ranked decreasingly according to their scores, and the links on the top are considered to be of the highest existence likelihoods. $S_{xy}$ is usually called similarity index in literatures.

To test the algorithm's accuracy, we randomly divide the link set $E$ into a training set $E_{T}$ and a test set $E_{V}$, so that, $E = E_{T}\cup E_{V}$ and $E_{T}\cap E_{V} =\phi$. In this study, $E_{T}$ and $E_{V}$ contain 90\% and 10\% links of $E$, respectively. We use precision to quantify the accuracy, which is the ratio of real missing links to those taken as predicted links, i.e., if the top $L$ links are considered as predicted links while $L_{r}$ of which appear in the test set, the precision is $L_r/L$.

In most weighted networks, the link weights actually measure pairwise affinities between nodes. That is, the larger the weight, the higher extent of relevance between corresponding nodes. When conducting link prediction, the likelihood score $S_{xy}$ is defined to measure the possibility of the existence of a link between nodes $x$ and $y$, which in essence reflects the affinity extent of nodes $x$ and $y$. Therefore, we can adjust the likelihood scores to predict link weights. Specifically, same as in link prediction, the link set $E$ is randomly divided into a training set $E_{T}$ and a test set $E_{V}$, thus the weighted adjacency matrices of the training and test sets are $W_{T}$ and $W_{V}$, respectively. Then the likelihood score for link prediction is in fact a function $S(W_{T})$ of the input matrix $W_{T}$. We need to define a weight prediction function $F(W_{T})$ so that the difference between $F(W_{T})$ and $W_{V}$ can be as small as possible.

We set $F(W_{T})=\lambda \cdot S(W_{T})$ where $\lambda$ is a scaling coefficient which can be identified by solving the following optimization problem:
\begin{equation}
\label{op function}
\min_{\lambda}\| \lambda \cdot S(W_{T})-W_{V}\|_{F}
\end{equation}
\begin{equation}\nonumber
\label{op function1}
\emph{s.t.}\quad
0<\lambda\leq M_V/M_{ST}
\end{equation}
where $M_{V}$ and $M_{ST}$ are maximum values of elements in the matrices $W_{V}$ and $S(W_{T})$, respectively, and $\|\cdot\|$  denotes the Frobenius norm, defined as the square root of the sum of the squares of the matrix's elements \cite{11}. We measure the accuracy of weight prediction by the Pearson correlation coefficient between the vectors of predicted and known link weights in the test set.

\section{Methods}
Systematical comparison suggested that the Common Neighbors (CN), Adamic-Adar (AA) and Resource Allocation (RA) indices perform best among lots of local similarity indices in unweighted networks\cite{1,12,13,14}. Thus we focus on these three measures whose definitions are as follows.\\
(i) CN index. The CN index simply counts the number of common neighbors between nodes $x$ and $y$ as:
\begin{equation}
\label{CN}
S_{xy}=|\Gamma(x)\cap \Gamma(y)|;
\end{equation}
where $\Gamma(x)$ is the set of neighbors of node $x$ and $|\cdot |$ denotes cardinality of the set.\\
(ii) AA index\cite{15}. This index depresses the contribution of the high-degree common neighbors by assigning larger contribution to less-connected neighbors:
\begin{equation}
\label{AA}
S_{xy}=\sum_{z\in \Gamma(x)\cap \Gamma(y)}\frac{1}{\log{k_{z}}};
\end{equation}
where $k_{z}$ is the degree of node $z$.\\
(iii) RA index\cite{12}. Similar to AA index, RA index also punishes the high-degree common neighbors, but to a higher extent, as
\begin{equation}
\label{RA}
S_{xy}=\sum_{z\in\Gamma(x)\cap\Gamma(y)}\frac{1}{k_{z}}.
\end{equation}

Previous studies extended similarity indices from unweighted networks to weighted networks by introducing the sum of weights of the two links $(z,x)$ and $(z,y)$, where $z$ runs over all common neighbors of nodes $x$ and $y$, as \cite{5,9}:\\
(i) Weighted CN index (WCN):
\begin{equation}
\label{wcn}
S_{xy}=\sum_{z\in\Gamma(x)\cap\Gamma(y)}(W_{xz}+W_{zy});
\end{equation}
(ii) Weighted AA index (WAA):
\begin{equation}
\label{WAA}
S_{xy}=\sum_{z\in\Gamma(x)\cap\Gamma(y)}\frac{W_{xz}+W_{yz}}{\log{(1+S_{z})}},
\end{equation}
(iii) Weighted RA index (WRA):
\begin{equation}
\label{WRA}
S_{xy}=\sum_{z\in\Gamma(x)\cap\Gamma(y)}\frac{W_{xz}+W_{yz}}{S_{z}}.
\end{equation}
Here, $S_z$ denotes the strength of node $z$, namely the sum of weights of associated links, as
\begin{equation}
S_z=\sum_{z'\in\Gamma(z)} W_{zz'}.
\end{equation}

In operations research, the most reliable route problem on a communication network requires to select the most reliable route to transmit data packages from a source node to a destination node. That is, one needs to find a route that links these two nodes and maximizes the probability that a package can reach the destination but not be corrupted in a non repairable fashion on the route. In this case, the communication network is represented as a weighted network, in which the weight of a link is its reliability, i.e., the probability that this link is safe for data transmit. Thus the reliability of a route is the product of weights of links along this route \cite{10}. As we have mentioned, weight in most weighted networks actually measures pairwise affinity between nodes, which is usually strongly related to reliability. For example, weight of a collaboration network is the number of co-authorized publications between two scientists, which is statistically correlated to the possibility that these two scientists will collaborate in the future \cite{16}. In a protein-protein interaction network, weight is the confidence score of the interaction, which represents the probability that the interaction occurs \cite{17}. In a collaborative network of e-commerce users, weight characterizes the co-purchases between two users \cite{18}. Thus we think it is reasonable to measure the affinity of a pair of unconnected nodes by the product of weights of paths connecting them. Therefore, we define the so-called reliable-route-based weighted similarity indices as follows:\\
(i) Reliable-route-based weighted CN index (rWCN):
\begin{equation}
\label{rwcn}
S_{xy}=\sum_{z\in\Gamma(x)\cap\Gamma(y)}(W_{xz}\cdot W_{zy})
\end{equation}
(ii) Reliable-route-based weighted AA index (rWAA):
\begin{equation}
\label{rWAA}
S_{xy}=\sum_{z\in\Gamma(x)\cap\Gamma(y)}\frac{W_{xz}\cdot W_{yz}}{\log{(1+S_{z})}}
\end{equation}
(iii) Reliable-route-based weighted RA index (rWRA):
\begin{equation}
\label{rWRA}
S_{xy}=\sum_{z\in\Gamma(x)\cap\Gamma(y)}\frac{W_{xz}\cdot W_{yz}}{S_{z}}
\end{equation}

Notice that, the reliable-route-based indices are borrowed from the definition of route reliability in the most reliable route problem in communication networks, where the edge weight of the communication network is the probability that this link is in good condition, which is no more than 1. Therefore, for networks whose weights do not lie in the range [0,1], before calculating reliable-route-based weighted similarity indices, we first map their weights to (0,1) through
\begin{equation}
\label{transform}
w'=e^{-\frac{1}{w}},
\end{equation}
where $w$ and $w'$ denote the original and regulated weights, respectively. Since the function (12) is a one-to-one mapping, it is easy to extract the original weight $w$ from the weight $w'$.

\begin{table*}[!ht]
\caption{Link prediction accuracy measured by precision on the networks under study. \label{precision}}
\centering
{\begin{tabular}{c|ccccccccccc}
\hline
Networks &hsaPPI &Cel &Geom &String &Corum &String\_Corum\\
\hline
$CN$     &0.207 &0.123 &0.196 &0.171 &0.626 &0.249\\
$WCN$    &0.207 &0.131 &0.177 &0.186 &0.631 &0.274\\
$rWCN$   &0.206 &0.13 &0.163 &0.187 &0.67 &0.27\\
$AA$     &0.217 &0.127 &0.391 &0.18 &0.63 &0.263\\
$WAA$    &0.217 &0.134 &0.391 &0.197 &0.674 &0.285\\
$rWAA$   &0.216 &\textbf{0.137} &0.352 &0.198 &0.703 &0.28\\
$RA$     &0.221 &0.124 &0.485  &0.221 &0.845 &0.304\\
$WRA$    &0.223 &0.13 &\textbf{0.493} &0.240 &\textbf{0.895} &\textbf{0.323}\\
$rWRA$   &\textbf{0.23} &0.13 &0.479 &\textbf{0.244} &0.891 &0.315\\
\hline
\end{tabular}}
\end{table*}

\begin{table*}[!ht]
\caption{Weight prediction accuracy measured by Pearson correlation coefficient on the networks under study.\label{pcc}}
\centering
{\begin{tabular}{ccccccccccc}
\hline
Networks &hsaPPI &Cel &Geom &String &Corum &String\_Corum\\
\hline
$CN$     &0.283	&0.178	&0.246	&0.189	&0.668 &0.242\\
$WCN$     &0.292	&0.183	&0.239	&0.251	&0.710 &0.299\\
$rWCN$    &0.300	&0.181	&0.235	&0.348	&0.750 &0.386\\
$AA$      &0.286	&0.194	&0.347	&0.201	&0.699	&0.251\\
$WAA$     &0.294	&0.205	&0.227	&0.078	&0.749 &0.309\\
$rWAA$    &\textbf{0.305}	&0.206	&0.249	&0.108	&0.786 &0.400\\
$RA$      &0.236	&0.209	&0.431	&0.261	&0.795 &0.295\\
$WRA$     &0.240	&0.214	&0.446	&0.292	&0.832 &0.339\\
$rWRA$    &0.256	&\textbf{0.216}	&\textbf{0.454}	&\textbf{0.391}	&\textbf{0.858}	&\textbf{0.438}\\
\hline
\end{tabular}}
\end{table*}

\section{Results}
Six real weighted networks are used to test our method. (i) hsaPPI: a high-confidence protein-protein interaction network of human beings constructed from the experimental biochemical co-fractionation data with overlap information derived from curated public databases and literature searches, in which the weight denotes the interaction confidence level \cite{20}. (ii) Cel: the updated version of the neural network of C.elegans, in which nodes are neurons, links are synaptic contacts between neurons, and the weight represents numbers of synapses between the neuron pair \cite{19}. (iii) Geom: the collaboration network covering scientific collaborations between scientists in computational geometry till February 2002, in which the link weight corresponds to the number of co-authorized publications between two scientists (can be downloaded from the \emph{Pajek Datasets} or the \emph{Computational Geometry Database}). (iv) String: weighted human gene-association network constructed from the database STRING \cite{22} (Search Tool for the Retrieval of Interacting Genes/Proteins). STRING integrates both physical interactions and functional associations from numerous sources and provides each link with a probabilistic confidence score. The version 9.05 of STRING was downloaded and used in this study. (v) Corum: a protein-protein interaction network of component proteins in human protein complexes collected by the database CORUM (Comprehensive Resource of Mammalian protein complexes) \cite{21}. We downloaded the database CORUM in June of 2013, whose core data includes 1343 complexes and 2314 component proteins. In this network, two proteins are linked if they appear in the same complex while the weight of each link is given by the number of shared complexes. (vi) String\_Corum: a sub-network of String constructed by extracting the proteins in CORUM and their links from the network String. See Table \ref{topology} for the basic topological measures of these networks. In Cel, Geom and Corum, weights stand for the numbers of synapses, co-authors and shared complexes, respectively. As mentioned above, we will first transform the weight $w$ in these three networks to area (0,1) by Eq. (12).

For each of the six networks, we randomly split its links into training and test set, which contain 90\% and 10\% of the links, respectively. Accuracy of link prediction is measured by precision, while accuracy of weight prediction is assessed by the Pearson correlation coefficient between the vectors of predicted and known link weights in the test set. When calculating precision metric, we set $L$ as the size of the test set. Repeating this process for 100 times, we obtained the average accuracy measures for link and weight prediction as listed in Table \ref{precision} and Table \ref{pcc}. For each network, the bold number in the row highlights the best prediction result.

Tables \ref{precision} and Table \ref{pcc} show that the algorithms of RA series perform overall best for both link and weight predictions, which is in consistent with earlier studies for link prediction \cite{1}. For link prediction, the only exception is Cel, whose highest precision is obtained by rWAA, while for weight prediction, the only exception is hsaPPI, whose best correlation coefficient comes from rWAA. In particular,  rWRA significantly outperforms others in weight prediction. 

It can be seen that the network Corum gets remarkably much higher prediction accuracy than the other networks, which may be because of that the network contains a plenty of cliques, as indicated by its large clustering coefficients for both unweighted and weighted versions (see Table 1), respectively defined as \cite{23,24}:
\begin{equation}
C(i)=\frac{\sum_{jk}a_{ij}a_{jk}a_{ki}}{\sum{_{jk}a_{ij}a_{ki}}},
\end{equation}
where $a_{ij}$ equals to 1 when there is a link between node $i$ and node $j$, else $a_{ij}$ is zero. And
\begin{equation}
C_{w}(i)=\frac{\sum_{jk}w_{ij}w_{jk}w_{ki}}{\max_{ij}w_{ij}\sum{_{jk}w_{ij}w_{ki}}},
\end{equation}
where $w_{ij}$ represents the weight of link $(i,j)$. The clustering coefficient of a network is the average clustering coefficient over all nodes whose degree is bigger than 1. Since the current local indices only takes into account common neighborhoods of two nodes, it is very straightforward to infer that the larger the clustering coefficient, the better the prediction. Indeed, the highest precision accuracies, corresponding to the bold numbers in Table 2 and Table 3, exhibit a strongly positive correlation with the clustering coefficients, as indicated by the high Pearson correlation coefficients reported in Table 4. Furthermore, one can also infer (not solid yet) that the weighted clustering coefficient gives a better characterization than the unweighted version, as indicated by the larger correlation coefficients.

\begin{table}[!ht]
\caption{Pearson correlations between clustering coefficients and the highest prediction accuracies corresponding to bold numbers in Table 2 and Table 3.\label{ccp}}
\centering
{\begin{tabular}{ccc}
\hline
 &Link Prediction & Weight Prediction \\
\hline
$C$     &0.866	&0.773\\
$C_w$     &0.940	&0.863\\
\hline
\end{tabular}}
\end{table}

\section{Conclusion and Discussion}
This work aims to predict missing links and their weights based on local information. A reliable-route-based method is proposed to generalize similarity indices defined from unweighted to weighted networks. Extensive experiments on real networks show that the series of resource allocation indices performs overall best in both link prediction and weight prediction. In addition, we show that the larger the clustering coefficient is, the higher the prediction accuracy is.

The prediction accuracy is also affected by the network background. For example, there are four networks whose nodes are human proteins. The network hsaPPI only includes high-confidence physical interactions between proteins, which is thus the most sparse network among the four. The network String comprises functionally associated pairs, including physical interactions, co-expression, co-localization, forming complex, and participating same biological process. In fact, the database STRING come from experiments, curated databases, literature mining, and computational prediction. Therefore, String is the most noisy one and its links are built up by different organizing principles, which is usually not easy for link prediction algorithms. That may be the reason why prediction accuracies for hsaPPI and String are relatively poor. In comparison, the network Corum was constructed to represent theoretical links between component proteins of complexes, who is the most regularly organized and thus is most easily to be explained and predicted.

\acknowledgments
This work was partially supported by the National Natural Science Foundation of China under Grant Nos. 61372194, 81260672 and 11222543. T.Z. acknowledge the program for New Century Excellent Talents in University under Grant No. NCET-11-0070.

\end{document}